\documentclass[useAMS,referee,usenatbib]{biom}
\usepackage{natbib}
\usepackage{amsmath}
\usepackage{amssymb}
\usepackage{bbm}
\usepackage{booktabs}
\usepackage{multirow}
\usepackage{array}
\usepackage{rotating}
\usepackage{graphicx}
\usepackage{tcolorbox}
\usepackage{subfig}

\def\bSig\mathbf{\Sigma}

\title[]{Bayesian Multiple Index Models for Environmental Mixtures}

\author{Glen McGee$^{1,*}$\email{glen.mcgee@uwaterloo.ca}, 
	Ander Wilson$^{2}$,
	Thomas F. Webster$^3$, and 
	Brent A. Coull$^{4}$ \\
	$^{1}$Department of Statistics and Actuarial Science, University of Waterloo, Waterloo, ON, Canada\\
	$^{2}$Department of Statistics, Colorado State University, CO, U.S.A. \\
	$^{3}$Department of Environmental Health, Boston University, Boston, MA, U.S.A.\\
$^{4}$Department of Biostatistics, Harvard T. H. Chan School of Public Health, Boston, MA, U.S.A.  } 

\begin{document}
	
	
	
	
	
	\pagerange{\pageref{firstpage}--\pageref{lastpage}} 
	\volume{000}
	\pubyear{0000}
	\artmonth{000}
	
	
	\doi{000}
	
	
	\label{firstpage}
	
	
	\begin{abstract}
An important goal of environmental health research is to assess the risk posed by mixtures of environmental exposures. Two popular classes of models for mixtures analyses are response-surface methods and exposure-index methods. Response-surface methods estimate high-dimensional surfaces and are thus highly flexible but difficult to interpret. In contrast, exposure-index methods decompose coefficients from a linear model into an overall mixture effect and individual index weights; these models yield easily interpretable effect estimates and efficient inferences when model assumptions hold, but, like most parsimonious models, incur bias when these assumptions do not hold. In this paper we propose a Bayesian multiple index model framework that combines the strengths of each, allowing for non-linear and non-additive relationships between exposure indices and a health outcome, while reducing the dimensionality of the exposure vector and estimating index weights with variable selection. This framework contains response-surface and exposure-index models as special cases, thereby unifying the two analysis strategies. This unification increases the range of models possible for analyzing environmental mixtures and health, allowing one to select an appropriate analysis from a spectrum of models varying in flexibility and interpretability. In an analysis of the association between telomere length and 18 organic pollutants in the National Health and Nutrition Examination Survey (NHANES), the proposed approach fits the data as well as more complex response-surface methods and yields more interpretable results. 
	\end{abstract}

	%
	
	
	
	\maketitle
	
	
	%

	\section{Introduction}
	\label{s:intro}

An important goal of environmental health research is to quantify the risk posed by the  environmental exposures to which humans are exposed. Exposures of interest can include those from multiple domains, such as chemical stressors (e.g. metals, persistant organic pollutants, or particles) and non-chemical stressors such as psychosocial stress or diet.   Research has historically focused on the effects of individual environmental exposures, but this is unrealistic as humans are not exposed to a single pollutant in isolation.  As such, a priority of the National Institute of Environmental Health Sciences (NIEHS), and the environmental health sciences in general, is to understand  the impact of environmental \textit{mixtures} \citep{dominici2010protecting,carlin2013unraveling,taylor2016statistical}.  

Numerous statistical approaches have been proposed to estimate the health effects of mixtures. These have been thoroughly reviewed in the literature \citep{billionnet2012estimating,braun2016can,davalos2017current,hamra2018environmental,gibson2019overview,lazarevic2019statistical,gibson2019complex,tanner2020environmental}. When choosing an appropriate method, practitioners  need to identify the primary scientific question of interest \citep{gibson2019complex} and then choose the appropriate statistical method for that question. This choice often includes deciding between highly flexible methods that are often hard to interpret and more restrictive methods that yield interpretable results. Two of the most popular mixture models in environmental health research, exposure-response surface  methodology and index based methods, exemplify this trade off.

Exposure-response surface  methods, such as Bayesian kernel machine regression (BKMR; \citealp{bobb2015bayesian,bobb2018statistical}) or Gaussian process regression more generally \citep{williams2006gaussian,ferrari2019identifying}, estimate a multi-dimensional exposure-response surface non-parametrically. This class of models allows for estimation of  non-linear and non-additive relationships between exposures and response. As such, these models can describe a  broad array of exposure-response relationships. In the specific case of BKMR, interpretation  is primarily based on posterior analysis of the high-dimensional exposure-response surface and interpretation is highly reliant on visualization. When there is a moderate to large number of components in the mixture, succinct interpretation can be difficult due to the large number of main effect and interaction surfaces.  For example, in the analysis of 18 exposures considered in this paper, BKMR requires visual inspection of 18 main effect exposure-response functions and 306 pairwise interaction surface plots.

In contrast, exposure-index methods analyze the relationship between a response and a linear multi-exposure index. Among the most popular  approaches in the environmental health sciences are  weighted quantile sum regression (WQS; \citealp{carrico2015characterization,renzetti2019gwqs,colicino2019per}) and  quantile G-computation (QGC; \citealp{keil2020quantile}). These approaches first transform predictors to the quantile scale and then fit a linear model. The regression coefficients are then decomposed into (1) an ``overall index effect'' and (2) index weights indicating the relative contribution of each mixture component to the overall effect. This decomposition eases interpretation because there is a single parameter that characterizes the overall effect of the mixture, and a set of weights that reflects the relative importance of each exposure to this overall effect.  
In addition, model parsimony results in efficient inferences when model assumptions hold.  However, these methods are ultimately  instances of linear models and may not be sufficiently flexible to accurately model a more complex exposure-response relationship. 
Though previous work has incorporated higher order terms or interactions  \citep{keil2020quantile}, one must specify these parametrically, and moreover this comes at the cost of the convenient interpretation of index weights. 

From a statistical perspective, these index based methods can be viewed as a single index model with an assumed  linear association  between the outcome and an index formed as a weighted average of quantiled exposures.  More general formulations of the single index model (SIM) relax the linearity assumptions and  model the relationship between the response and the exposure index non-parametrically \citep{powell1989semiparametric,hardle1993optimal,naik2001single,hristache2001direct,yu2002penalized,xia2004goodness,lin2007identifiability,kong2007variable,xia2008multiple,wang2009spline,wang2010estimation}. In situations where there are natural groupings of components within a mixture, one can further allow for multiple indices and interactions among them in a multiple index model (MIM) framework (\citealp{stoker1986consistent,ichimura1991semiparametric,samarov1993exploring,hristache2001structure,horowitz2012semiparametric,xia2008multiple}). These approaches have received substantial attention in the econometric and statistical literature but have not enjoyed the same popularity in  environmental health: all of eight recent reviews of methods for analyzing multi-pollutant mixtures cited WQS/QGC and BKMR, but none considered MIMs \citep{taylor2016statistical,braun2016can,davalos2017current,hamra2018environmental,gibson2019overview,lazarevic2019statistical,gibson2019complex,tanner2020environmental}.  Recently \cite{wang2020singleindex} applied a frequentist SIM for exposure mixtures, but we are unaware of any application of MIMs to mixtures analysis. In terms of methodological innovation, while Bayesian methods for fitting SIMs have been developed \citep{antoniadis2004bayesian,park2005bayesian,choi2011gaussian,gramacy2012gaussian,alquier2013sparse,reich2011sufficient,poon2013bayesian}, we are not aware of Bayesian methods for MIMs.

In this paper, we propose a Bayesian multiple index model (BMIM), which provides a unified framework for estimating the association between exposure mixtures and a health outcome. Specifically, the BMIM represents a class of models that includes both response surface and index-based models. 
This class of models also encompasses a spectrum of models between these two extremes that vary in terms of model flexibility and interpretability, thus giving analysts greater control in selecting an appropriate method for any given data analysis.  

The proposed approach: (1) facilitates formal Bayesian inference on interpretable index weights; (2) allows for a non-linear relation between an index and the outcome; (3) permits non-additive interactions among multi-exposure indices;  and (4) incorporates Bayesian variable selection on mixture components. Moreover, we extend the BMIM framework to accommodate binary responses as well as cluster-correlated data.

In Section \ref{s:NHANES} we introduce the motivating case study based on data from the National Health and Nutrition Examination Survey (NHANES).  We briefly outline two of the most popular methods for mixtures analyses in the environmental health literature---QGC and BKMR---in Section \ref{s:existing}. In Section \ref{s:MIM} we describe the proposed Bayesian multiple index framework. We then compare methods via simulation study (Section \ref{s:simulations}) and in the NHANES case study (Section \ref{s:casestudy}). Finally, we conclude with a discussion in Section \ref{s:discussion}.

\section{NHANES Case Study: Telemere Length and a Mixture of Pollutants}
\label{s:NHANES}
We consider a case study of the association between a mixture of environmental pollutants and leukocyte telomere length (LTL), an important biomarker of cellular aging \citep{mitro2016cross}. To investigate this question, \cite{mitro2016cross}, and later \cite{gibson2019overview}, analyzed data from the 2001–2002 National Health and Nutrition Examination Survey (NHANES). From the original cohort of 11,039, both authors analyzed a subset of 1003 people who were at least 20 years of age and had complete data for a number of variables. 
 Following \cite{gibson2019overview}, we consider an exposure mixture including 18 persistent organic pollutants that have been grouped naturally into sets of pollutants hypothesized to act similarly: Group 1 includes eight non-dioxin-like polychlorinated biphenyls (PCBs 74, 99, 138, 153, 170, 180, 187, and 194); Group 2 includes two non-ortho PCBs (PCBs 126 and 169); Group 3 includes PCB 118, four furans, and three dioxins. The outcome of interest is log-LTL, which is thought to be susceptible to  environmental exposures \citep{mitro2016cross}. Further details on exposure and outcome measurement, as well as on inclusion criteria, have been reported in depth elsewhere \citep{mitro2016cross,gibson2019overview}.

In what follows, we analyze the same sample of $N$=1003 observations with the goal of characterizing the relationship between log-LTL and the mixture of pollutants as well as the individual pollutant contributions.

\vspace{-0.5cm}

\section{Existing Methods}
\label{s:existing}
First let $y_i$ be the outcome of interest (log-LTL), $\{x_{i1},\cdots,x_{iP}\}$ be the exposures ($P$=18 pollutants), and $\mathbf{z}_i$ be a vector of covariates for the $i^{th}$ observation ($i=1,\cdots,N$).  

\subsection{Quantile G-Computation (QGC)}
\label{s:qgcomp}


Exposure index methods such as WQS \citep{gennings2010identifying,carrico2015characterization} and QGC \citep{keil2020quantile} fit a linear model as follows:
\begin{align}
y_i&=\alpha_0 + \beta\left(b_1 q_{i1}+\cdots+b_P q_{iP}\right) +\mathbf{z_i}^T\boldsymbol{\gamma}+\epsilon_i, \label{eqn:wqs} 
\end{align}
where $q_{ip}$ represent pre-transformed versions of $x_{ip}$, scored as quantiles (e.g., 0,1,2,3 if $x_{ip}$ is in the  0-1st, 1-2nd, 2-3rd or 3-4th quartile). The parameter $\beta$ represents the linear association between the index and the outcome. Rather than constructing an index based on weights fixed \textit{a priori}, WQS and QGC regression  simultaneously estimate an overall index association and component weights. 

To identify both the weights and the overall effects some identifiability constraint is needed. For WQS, $\sum_j b_j =1$ and $b_j \geq 0~\forall j$ (which we do not pursue here).  QGC relaxes the positivity assumptions and allows for both positive and negative associations.  Ultimately, the QGC approach estimates a linear model and then decomposes the estimated coefficients into an overall effect beta and component weights $b_{p'}=\frac{\beta_{p'}}{\sum_p \beta_p}$ post-hoc.  When estimates are of opposite sign, QGC typically  reports positive weights ($b_{p'}^+=\beta_{p'}1(\beta_{p'}>0)/\sum_p \beta_{p}1(\beta_{p'}>0)$) and negative weights  ($b_{p'}^-=\beta_{p'}1(\beta_{p'}<0)/\sum_p \beta_{p}1(\beta_{p'}<0)$) separately (see the {\tt qgcomp} package; \citealp{keil2020quantile}). As such, each is interpreted as the ``proportion of the positive association,'' and analogously so for the negative weights \citep{keil2020quantile}.

\subsection{Bayesian Kernel Machine Regression (BKMR)}
\label{s:bkmr}
We consider BKMR as a popular response surface method for environmental mixtures. BKMR is a flexible approach to modelling mixtures that allows non-linear associations and non-additive interactions among exposures. The BKMR model is\label{ss:bmim}
\begin{align}
y_i&=h^P\left(x_{i1},\dots,x_{iP}\right) +\mathbf{z}_i^T\boldsymbol{\gamma}+\epsilon_i, \label{eqn:bkmr} 
\end{align}
where $h^P(\cdot): \mathbb{R}^P \to \mathbb{R}$ is an unknown and potentially non-linear function represented via a kernel. We assume $h^P(\cdot)$ exists in $\mathcal{H}_K$, defined by a positive semidefinite reproducing kernel $K:\mathbb{R}^P \times \mathbb{R}^P \to \mathbb{R}$.

The choice of kernel function $K(\cdot,\cdot)$ uniquely determines a set of basis functions \citep{cristianini2000introduction}. 
A common choice is the Gaussian kernel,
$K(\mathbf{x},\mathbf{x'})=exp\left[-\sum_{p=1}^P \rho_p (x_p-x'_p)^2\right],$ where  $\rho_p \geq 0$ are feature weights and $\mathbf{x}=(x_1,\dots,x_P)$. This corresponds to  radial basis functions \citep{liu2007semiparametric}.

Estimation can proceed based on an equivalence with a linear mixed effects model \citep{liu2007semiparametric}. Under a Kernel representation for $h^P(\cdot)$, model (\ref{eqn:bkmr}) can be  written as
\begin{align*}
y_i|h_i &\sim N(h_i+\mathbf{z}_i^T\boldsymbol{\gamma},\sigma^2), \\
(h_1,\cdots,h_N)^T&\sim N(\mathbf{0},\lambda^{-1}\sigma^2  \mathbf{K}),
\end{align*} 
where $\mathbf{K}$ is the kernel matrix with elements $\mathbf{K}_{ij}=K(\mathbf{x}_i,\mathbf{x}_j)$, and $\lambda>0$ is a tuning parameter that determines model complexity with small $\lambda$  favoring a more flexible model \citep{liu2007semiparametric}.  Estimation is based on the marginal likelihood of $\mathbf{y}=(y_1,\cdots,y_N)^T$ with respect to $\mathbf{h}=(h_1,\cdots,h_N)^T$:
\begin{align}
\mathbf{y}\sim N\left[\mathbf{Z}\boldsymbol{\gamma},\sigma^2(\mathbf{I}+\lambda^{-1} \mathbf{K})\right]. \label{eqn:likelihood}
\end{align} The model is completed by specifying priors for $\{\boldsymbol{\rho},\boldsymbol{\gamma},\sigma^2,\lambda\}$.  In our simulations and data analysis we adopted the default priors proposed by  \cite{bobb2015bayesian,bobb2018statistical}.

Component-wise variable selection is incorporated via spike and slab priors on the $\rho_p$. To identify effects of exposures that may be highly correlated, \cite{bobb2015bayesian} also introduced a hierarchical variable selection scheme. This permits only one pollutant among a group of  exposures to enter the model at a time, which may be  too restrictive in some situations. 
\vspace{-0.5cm}

\section{Proposed Approach: Bayesian Multiple Index Models}
\label{s:MIM}

\subsection{Model Framework}
\label{ss:mimmethods}
We propose a Bayesian multiple index model (BMIM) framework to combine the flexibility of response surface methods with the interpretability of more parsimonious index models.  Suppose $x_{i1},\cdots,x_{iP}$ can be partitioned into $M$ ($M\in \{1,\dots,P\}$) mutually exclusive groups denoted  $\mathbf{x}_{im}=(x_{im1},\cdots,x_{imL_m})^T$ for $m=1,\dots,M$. 
The proposed model can be written
\begin{align}
y_i&=h^M\left(\mathbf{x}_{i1}^T \boldsymbol{\theta_1},\cdots,\mathbf{x}_{iM}^T \boldsymbol{\theta}_M\right) +\mathbf{z}_i^T\boldsymbol{\gamma}+\epsilon_i, \label{eqn:model1} 
\end{align}
where $\boldsymbol{\theta}_m$ are $L_m$-vectors of index weights subject to the identifiability constraints:
$
\boldsymbol{1_{L_m}}^T\boldsymbol{\theta}_m\geq 0$, where $\boldsymbol{1_{L_m}}$ is the unit vector of length $L_m$, and  $\boldsymbol{\theta}_m^T\boldsymbol{\theta}_m =1$ for $m=1,\cdots,M.$ Contrast these  constraints with those of the linear index models: like QGC, this allows for positive and negative associations, but rather than summing to 1 the weights have norm 1. 

The key notion is that, in contrast to BKMR, one need only estimate an exposure-response surface of dimension $M$$\leq$$P$, which may remain small even when $P$ is large.  Moreover, within the $m^{th}$ index, we can interpret the contributions of each exposure via index weights $\boldsymbol{\theta}_m$.

We again employ kernel function $K(\cdot,\cdot)$ that is now a function of a vector of the $M$ indices. 
That is,  the Gaussian kernel can be written as 
\begin{align}
K(\mathbf{E},\mathbf{E'})&=exp\left[-\sum_{m=1}^M \rho_m \{(\mathbf{x}_{m}- \mathbf{x^{'}}_{m})^T \boldsymbol{\theta}_m\}^2\right],\label{eqn:kern_gauss_BSMIM} 
\end{align}
for $\mathbf{E}=(\mathbf{x}_{1}^T \boldsymbol{\theta}_1,\cdots,\mathbf{x}_{M}^T \boldsymbol{\theta}_M)$ and $\mathbf{E'}=(\mathbf{x^{'}}_{1}^T \boldsymbol{\theta}_1,\cdots,\mathbf{x^{'}}_{M}^T \boldsymbol{\theta}_M)$.

To reduce the computational burden of sampling a vector from this constrained space in our MCMC algorithm, we reparameterize $\boldsymbol{\theta^*}_m=\rho_m^{1/2}\boldsymbol{\theta}_m$ as in \cite{wilson2019kernel}. The kernel in (\ref{eqn:kern_gauss_BSMIM}) is then:
\begin{align}
K(\mathbf{E},\mathbf{E'})&=exp\left[-\sum_{m=1}^M  \{(\mathbf{x}_{m}- \mathbf{x^{'}}_{m})^T \boldsymbol{\theta^*}_m\}^2\right].\label{eqn:kern_gaussstar} 
\end{align}
The weights $\boldsymbol{\theta}_m$ can be fully identified and recovered from the posterior sample of $\boldsymbol{\theta^*}_m$. Specifically, we have $\rho_m=||\boldsymbol{\theta^*}_m||^{2}={\boldsymbol{\theta^*}_m}^T\boldsymbol{\theta^*}_m$ and $\boldsymbol{\theta}_m=||\boldsymbol{\theta^*}_m||^{-1}\boldsymbol{\theta^*}_m$.

\subsection{Prior Specification and Posterior Inference}
\label{ss:priors}
We specify a prior directly for the unconstrained $\boldsymbol{\theta^*}_m$. In particular, we allow for variable selection on the exposure component weights via spike and slab priors:
\begin{align*}
P({\theta^*_{ml}}|\delta_{ml})&=  \delta_{ml} N(0, \sigma_{\theta}^2) + (1-\delta_{ml})P_0, ~~\text{for $l=1,\cdots,L_m$  s.t. } \mathbbm{1}(\mathbf{1}_{L_m}^T\boldsymbol{\theta^*}_m\geq 0),\\
P(\delta_{ml})&= Bernoulli(\pi), \\
P(\pi)&= Beta(a_0,b_0),
\end{align*}
where $P_0$ is a point mass at zero. Note that this applies selection at the component level, but can exert selection on an index when none of its components are selected into the model. Although we impose equal shrinkage at the \textit{component} level (i.e., each component has equal prior inclusion probability), one could entertain other specifications to, say, apply equal shrinkage to each \textit{index}.

To complete the model specification, we adopt a flat prior for  $\boldsymbol{\gamma} $, $\sigma^{-2}\sim \text{Gamma}(a_\sigma,b_\sigma)$ and $\lambda^{-1}\sim \text{Gamma}(a_{\lambda},b_{\lambda}) $,
and we set the hyperparameters to $a_\sigma=0.001,b_\sigma=0.001, a_\lambda=1, b_\lambda=0.1, a_0=1$, and  $b_0=1$.

Inference  follows from the marginal likelihood of $\mathbf{y}$, which takes the same form as (\ref{eqn:likelihood}) with the relevant kernel matrix. After marginalizing over $\pi$, the posterior can be decomposed as:
%
\begin{align*}
P(\boldsymbol{\theta^*}_1,\cdots,\boldsymbol{\theta^*}_M,\boldsymbol{\delta},\boldsymbol{\gamma} ,\sigma^{-2},\lambda^{-1}|\mathbf{y}) &=P(\boldsymbol{\gamma}|\boldsymbol{\theta^*}_1,\cdots,\boldsymbol{\theta^*}_M,\boldsymbol{\delta},\sigma^{-2},\lambda^{-1},\mathbf{y}) \\
&\times P(\sigma^{-2}|\boldsymbol{\theta^*}_1,\cdots,\boldsymbol{\theta^*}_M,\boldsymbol{\delta},\lambda^{-1},\mathbf{y}) \\
&\times P(\boldsymbol{\theta^*}_1,\cdots,\boldsymbol{\theta^*}_M,\boldsymbol{\delta},\lambda^{-1}|\mathbf{y}).
\end{align*}
We draw from the posterior via MCMC (see Appendix for details). To avoid sampling $\sigma^{-2}$ and $\gamma$ at every iteration, we integrate over them and draw from the marginal posterior $P(\boldsymbol{\theta^*}_1,\cdots,\boldsymbol{\theta^*}_M,\boldsymbol{\delta},\lambda^{-1}|\mathbf{y})$ 
 via modified Gibbs with Metropolis steps, iterating between drawing $\lambda^{-1}$ and drawing $(\boldsymbol{\theta^*}_1,\cdots,\boldsymbol{\theta^*}_M,\boldsymbol{\delta})$ jointly (as in \citealp{bobb2015bayesian}).  After thinning and applying burn-in, we  draw $\sigma^{-2}$ directly from a Gamma distribution and $\gamma$ from a normal distribution. Finally, we obtain posterior samples of the interpretable weights $\boldsymbol{\theta}_m$ by decomposing draws of $\boldsymbol{\theta^*}_m$ as described above.  

We summarize the posterior for the weights as follows. We first report the posterior inclusion probability (PIP) for an entire index---i.e. the posterior probability that $\rho_m$$\neq$0. When a whole index is selected out of the model ($\rho_m=0$), the component weights $\theta_{ml}$ are undefined. In this case, we subsequently describe  the conditional posterior for $\theta_{ml}|\rho\neq0$ via PIP, mean and 95\% credible intervals. To summarize component-wise variable selection, we recover marginal PIPs for the component weights by multipling the PIP for $\rho_m$ by the conditional PIP  for $\theta_{ml}$, as $P(\theta_{ml}\neq 0 )=P(\theta_{ml}\neq 0|\rho_m\neq 0 )P(\rho_m\neq 0 )$.\vspace{-0.105cm}

\subsection{Estimating the Exposure-Response Surface}
\label{ss:predict}
Posterior draws for $h_i$ at the observed exposure levels can be obtained following \cite{bobb2015bayesian}. Details can be found in the supplementary material. More often, we are interested in predicting $\mathbf{h}^{new}$ on a grid of $G$ new exposure levels. Let $\mathbf{E}^{new}=({\mathbf{E}^{new}_1}^T,\dots,{\mathbf{E}^{new}_G}^T)^T$ be the $G\times M$ matrix of new levels. Recall that $\mathbf{K}$ is the $N\times N$ matrix with elements $\mathbf{K}_{ij}=K(\mathbf{E}_i,\mathbf{E}_j)$. Let $\mathbf{K}^{nn}$ be the $G\times G$ kernel matrix for the new levels, with elements  $\mathbf{K}^{nn}_{ij}=K(\mathbf{E}^{new}_i,\mathbf{E}^{new}_j)$.  Similarly let $\mathbf{K}^{no}$ be the $G\times N$ kernel matrix comparing new exposure levels to observed ones, with elements $\mathbf{K}^{no}_{ij}=K(\mathbf{E}^{new}_i,\mathbf{E}_{j})$. The posterior predictive distribution is 
\begin{align*}
\mathbf{h}^{new}&\sim MVN\left[\lambda^{-1}\mathbf{K}^{no}(\mathbf{I}+\lambda^{-1}\mathbf{K})^{-1}(\mathbf{y}-\mathbf{Z}\boldsymbol{\gamma}),
\sigma^2\lambda^{-1}\{\mathbf{K}^{nn}-\lambda^{-1}\mathbf{K}^{no}(\mathbf{I}+\lambda^{-1}\mathbf{K})^{-1}{\mathbf{K}^{no}}^T \}\right].
\end{align*}

\subsection{Indexwise Curves and Other Associations of Interest}

As with BKMR, we can predict the response surface at arbitrary exposure levels. As such, BMIM estimation yields analogous component-wise response curves component-wise response curves by varying a single exposure along a grid and holding all others fixed.  This approach also allows one to report an ``overall mixture effect'' analogous to that of QGC by simultaneously increasing all exposures by a quantile, or an ``overall index effect'' by  simultaneously increasing  all exposures within an index. 

The BMIM structure lends itself  to  reporting index-wise response curves. Consider the $m^{th}$ index, and set $E^{new}_{gm'}$ to some fixed quantile of posterior means of $\left(\mathbf{x}_{i m'}^T \boldsymbol{\theta}_{m'}\right)$ for each $m'\neq m$ and $g=1,\dots,G$. Then set  $(E^{new}_{1m},\dots,E^{new}_{Gm})^T$ (i.e., the $m^{th}$ column of $\mathbf{E}^{new}$) to a grid of constants---e.g., equally spaced values between 5th and 95th percentiles across $N$ observations of the posterior means for each ${\mathbf{x}_{im}}^T \boldsymbol{\theta}_m$. This is a convenient choice in that it naturally captures how exposures vary jointly. In contrast, increasing exposures simultaneously by a quantile would not necessarily capture their joint variability unless exposures are highly correlated. Note that this implicitly takes the weights as fixed and thus isolates uncertainty in the shape of the index-response curve. That is, it ignores uncertainty in the weights that comprise the index of interest. This is akin to making inferences about $\beta$ in model (\ref{eqn:wqs}), or constructing indices via (fixed) toxic equivalency factors as is common in toxicology (e.g., \citealp{mitro2016cross}). 

The parsimonious set of index-wise exposure-response functions simplifies the interpretation of a fitted model, as one can plot $M$ index-wise curves rather than $P$ component-wise curves. The contrast is even starker for interactions, as one could present $M\times(M-1)$ two-way index-wise interaction plots rather than $P\times(P-1)$. In the NHANES example, this corresponds to 306 component-wise interaction plots under BKMR and only 6 index-wise plots under a MIM with 3 indices. Nevertheless, one could  still extract the same component-wise curves while reflecting uncertainty associated with the estimation of the weights

\subsection{Relation to Existing Methods and a Spectrum of Models}
\label{ss:connections}

The proposed framework contains two special cases worth highlighting.  First, by setting $M=1$, the BMIM reduces to a single index model. In particular, if one were to adopt a polynomial kernel:
$K(\mathbf{E},\mathbf{E'})=\left[1+\sum_{m=1}^M \rho_m (\mathbf{x}_{m}^T\boldsymbol{\theta}_m)(\mathbf{x^{'}}_{m}^T\boldsymbol{\theta}_m )\right]^d,$ 
of degree $d=1$ (and pre-transform exposures accordingly) one could recover the index model in (\ref{eqn:wqs})---with the  benefit of uncertainty quantification on the weights $\theta_{ml}$. Even in more flexible single-index specifications, say with a higher order polynomial or a Gaussian kernel, one can still estimate well-defined and interpretable weights. Hence, the  class of index models is contained in BMIM, including standard linear models and more flexible models with a nonlinear association.

At the other extreme,  setting $M$=$P$---that is, 
a BMIM model containing P single-exposure indices---corresponds to the ususal BKMR. This shows that the two common approaches to analyzing environmental mixtures in fact lie on a opposite ends of a spectrum of models that vary in their flexibility. Rather than forcing analysts to choose between one of these two extremes, the BMIM framework permits one to specify the level of flexibility and interpretability most appropriate for a given analysis.  

\subsection{Software and Extensions}
\label{ss:extensions}
In addition to the methods described here---including both Gaussian and polynomial kernels---we have extended the BMIM to several common scenarios. 
When outcomes are cluster-correlated, the BMIM can incorporate random intercepts. It can also accommodate binary outcomes via a probit-latent variable formulation. Details of these extensions can be found in the supplemental material. Associated software is available at {\tt github.com/glenmcgee/BMIM}.

\section{Simulations}
\label{s:simulations}

\subsection{Setup}
We conducted a series of simulations to compare the behaviour of Bayesian single- and multiple-index models to the more flexible BKMR and the parsimonious QGC. Using the real exposures ($P$=18) and covariates in the NHANES sample, we generated outcomes from:
\begin{align*}
y_i&\sim N(h_i+\mathbf{z}_i^T\boldsymbol{\gamma},\sigma^2),
\end{align*}
where $\mathbf{z}_i$ included age (standardized), age$^2$, male (0,1), and indicators of BMI (25–-29.9; 30+). We set $\gamma=(-0.43,0.00,-0.25,0.12,0.08)^T$ to loosely reflect effect sizes in the data application, and we set $\sigma=0.5$. Finally, we assumed two different structures for $h_i=h^*(\mathbf{x}_i) $:
\begin{enumerate}
	\item[(A)] Scenario A follows from a single index model. First generate ${x}^*_i=\mathbf{x}_i^T\mathbf{w}$ where $\mathbf{w}=(\mathbf{w}_1^T,(0,0)^T,\mathbf{w}_3^T)^T$ such that $\mathbf{w}_1=(8,7,6,5,4 ,3,2,1)^T$ and $\mathbf{w}_3=-2\times \mathbf{1}_{8}$, standardized to have mean 0 and standard deviation 1. Then $h_i=h_A(x^*_i)$, where $h_{A}(\cdot)$ is S-shaped.
	\item[(B)] Scenario B follows from a multiple index model. Generate $\mathbf{x}_1^*=(x_1,\dots,x_8)\mathbf{w}_1$, $\mathbf{x}_2^*=(x_{9},x_{10})\mathbf{w}_2$ and $\mathbf{x}_3^*=(x_{11},\dots,x_{18})\mathbf{w}_3$, where $\mathbf{w}_2=(1,1)^T$, standardized as above, and $\mathbf{w}_1$ and $\mathbf{w}_3$ are as defined above. Then
	\begin{align*}
	h_i&=h_{B1}^*(x_{1i}^*)+h_{B2}^*(x_{2i}^*)+h_{B3}^*(x_{3i}^*)+0.5h_{B1}^*(x_{1i}^*)\times h_{B3}^*(x_{3i}^*) 
	\end{align*}
	where  $h_{B1}$ is a unimodal function, $h_{B2}$ is flat (null), and $h_{B3}$ is sigmoidal.
\end{enumerate}
Details on the exposure response curves $\{h_{A},h_{B1},h_{B2},h_{B3}\}$ can be found in the Appendix. 

In each setting, we generated 100 datasets  of 500 observations. We then split each sample into a training set of $N=300$ observations, and held out the remaining $200$ observations on which to evaluate model fit.  To each training set, we then fit QGC (with $q$=10, i.e. deciles), a Bayesian single index model (BSIM), a 3-index BMIM using the same indices as above (BMIM-3), full BKMR, and BKMR using hierarchical variable selection using the three groups of exposures (BKMR-H), each with Gaussian kernel.

Intuitively, in scenario (A), we would expect the BSIM to perform best as it is correctly specified (with the shape of the response curve unknown). We expect BMIM (with $M=3$) and BKMR to be flexible enough to still perform well here, although with more variability due to less model structure. In scenario (B), the BSIM is mis-specified and hence should perform poorly. Here BMIM is correctly specified (with unknown response curve) and should perform better than the more flexible BKMR due to its increased structure. QGC is misspecified in both scenarios, as it assumes linearity on the quantile scale.

We evaluated model fit via mean squared error (MSE) for the unknown $h_i$ in the test set, as well as MSE for the outcomes in the test set and in 4-fold cross validation (CV). For MSE,  we report the mean as well as the standard deviation across datasets in order to contextualize differences between models. We  also report 95\%  interval coverage (Cvg) and average standard errors (SE) for $h_i$ in the test set.


\subsection{Simulation Results}
Under the single index data-generating mechanism (A), the BSIM had the lowest MSE. The next lowest MSE was for BMIM-3 followed by BKMR.   Under the three-index data-generating mechanism (B),  BMIM-3 had the lowest MSE followed by BKMR. The incorrectly specified BSIM performed worse than both. In both scenarios, QGC exhibited very high MSE relative to the other approaches, because it was misspecified. As described previously, one could incorporate higher order terms into the QGC approach, but we found that incorporating quadratic terms for all exposures still did not perform  well  (see supplementary material). Cross validation MSE followed the same pattern, suggesting this can be used for model selection in practice.

Results for interval coverage  behaved similarly,  with the simplest correctly specified model achieving near nominal coverage.  Interestingly, under the single-index scenario (A), the more flexible alternatives (BMIM-3 and BKMR) had progressively lower coverage, potentially because of bias due to applying more shrinkage; nevertheless both greatly outperformed QGC. In scenario (B), BMIM-3 and BKMR both achieved near nominal coverage whereas the misspecified BSIM and QGC had very low coverage. Naturally, the more structured  BSIM had smaller SEs than BMIM-3, which had lower SEs than BKMR in  (A) and (B).  

Interestingly, when we repeated the simulations in low-signal settings (setting $\sigma$=1.0 or 2.0), the simpler models were no longer the clear winners. Instead, the BSIM, BMIM-3 and BKMR all performed about the same in terms of MSE, interval coverage, and even average SEs.  Indeed, cross validation MSEs were effectively equal in low-signal settings. These models fit about equally well, and it would be difficult to choose one that best fits the data. Facing such a setting in practice, one would likely favor the model with simpler interpretations.

Note that we report here results for BKMR using the same priors for $\rho_m$ as in the BMIM and BSIM. In the supplementary material we report results for BKMR using the default priors for $\rho_m$ from the {\tt bkmr} package \cite{bobb2018statistical}, which performed very similarly to our implementation when $\sigma$ was small but  worse when $\sigma$ was high. BKMR with hierarchical variable selection generally performed worse in terms of MSE and interval coverage but  achieved the lowest average SEs (see supplementary material), because it imposes the most shrinkage by limiting a maximum of one exposure per group to feature in the model at a time. Therefore it is misspecified for the data generation mechanisms considered here. 

We also repeated scenario A with a linear response curve, and the results are similar to those reported here (see Appendix). Even here, QGC remains misspecified, as it assumes linearity on the quantile scale. As such it still performed poorly relative to the kernel approaches. Unsurprisingly, BSIM with a polynomial kernel of degree one performed best.

\begin{table}[htbp!]
	\centering
	\caption{Simulation Results across 100 datasets. 
		The table shows mean squared error (MSE), average standard errors ($\overline{\text{SE}}$) and 95\% interval coverage (Cvg) on the estimated $h$ function in the test dataset, as well as MSE for $y$ evaluated on a test dataset (MSE($y$)) and via cross validation (CV-MSE($y$)).
		BKMR-H is BKMR with hierarchical variable selection.}
	\begin{tabular}{rrrccccrrrr}
		\toprule
		&&& \multicolumn{2}{c}{MSE($h$)} & $\overline{\text{SE}}$($h$) & Cvg($h$) & \multicolumn{2}{c}{MSE($y$)} & \multicolumn{2}{c}{CV-MSE($y$)} \\
		\cmidrule{4-5}   \cmidrule{8-9} \cmidrule{10-11}
		$\sigma$&$M$& Model& Mean & SD & ~~ & ~~ & Mean & SD & Mean & SD \\ 
		\midrule
		0.5&1&QGC & 0.150 & 0.015 & 0.19 & 0.63 & 0.41 & 0.04 & 0.44 & 0.04 \\ 
		&&BSIM & 0.031 & 0.010 & 0.17 & 0.94 & 0.28 & 0.03 & 0.30 & 0.03 \\ 
		&&BMIM-3 & 0.051 & 0.011 & 0.21 & 0.91 & 0.30 & 0.03 & 0.34 & 0.03 \\ 
		&&BKMR & 0.076 & 0.012 & 0.22 & 0.83 & 0.32 & 0.04 & 0.37 & 0.03 \\ 
		&&BKMR-H  & 0.085 & 0.014 & 0.14 & 0.60 & 0.33 & 0.04 & 0.38 & 0.03 \\ 
		\\[-1.8ex]   
		&3&QGC & 0.188 & 0.024 & 0.19 & 0.61 & 0.41 & 0.04 & 0.42 & 0.04 \\ 
		&&BSIM & 0.122 & 0.027 & 0.17 & 0.54 & 0.34 & 0.04 & 0.35 & 0.04 \\ 
		&&BMIM-3 & 0.035 & 0.012 & 0.19 & 0.95 & 0.28 & 0.03 & 0.29 & 0.03 \\ 
		&&BKMR & 0.048 & 0.013 & 0.23 & 0.94 & 0.29 & 0.03 & 0.30 & 0.03 \\ 
		&&BKMR-H  & 0.085 & 0.021 & 0.14 & 0.62 & 0.33 & 0.04 & 0.35 & 0.03 \\ 
		\\[-1.8ex]   
	\midrule
	\\[-1.8ex]    
		1.0&1&QGC & 0.226 & 0.048 & 0.33 & 0.81 & 1.21 & 0.13 & 1.27 & 0.11 \\ 
		&&BSIM & 0.086 & 0.038 & 0.30 & 0.94 & 1.07 & 0.12 & 1.13 & 0.10 \\ 
		&&BMIM-3 & 0.113 & 0.033 & 0.31 & 0.90 & 1.10 & 0.12 & 1.17 & 0.10 \\ 
		&&BKMR& 0.125 & 0.034 & 0.30 & 0.87 & 1.11 & 0.12 & 1.19 & 0.10 \\ 
		&&BKMR-H  & 0.142 & 0.035 & 0.22 & 0.65 & 1.13 & 0.13 & 1.17 & 0.09 \\ 
		\\[-1.8ex]   
		&3&QGC & 0.263 & 0.059 & 0.33 & 0.77 & 1.21 & 0.12 & 1.27 & 0.12 \\ 
		&&BSIM & 0.157 & 0.042 & 0.30 & 0.81 & 1.13 & 0.11 & 1.15 & 0.10 \\ 
		&&BMIM-3 & 0.090 & 0.035 & 0.33 & 0.95 & 1.07 & 0.11 & 1.11 & 0.11 \\ 
		&&BKMR & 0.094 & 0.035 & 0.34 & 0.95 & 1.08 & 0.11 & 1.11 & 0.10 \\ 
		&&BKMR-H  & 0.141 & 0.046 & 0.23 & 0.71 & 1.12 & 0.12 & 1.15 & 0.10 \\ 
		\\[-1.8ex]   
		\midrule
		\\[-1.8ex]  
		2.0&1&QGC & 0.528 & 0.177 & 0.62 & 0.91 & 4.43 & 0.47 & 4.63 & 0.42 \\  
		&&BSIM & 0.227 & 0.113 & 0.48 & 0.92 & 4.17 & 0.45 & 4.28 & 0.36 \\ 
		&&BMIM-3 & 0.229 & 0.103 & 0.48 & 0.92 & 4.18 & 0.44 & 4.33 & 0.37 \\ 
		&&BKMR & 0.232 & 0.104 & 0.48 & 0.92 & 4.18 & 0.45 & 4.34 & 0.34 \\ 
		&&BKMR-H  & 0.291 & 0.120 & 0.37 & 0.75 & 4.24 & 0.46 & 4.29 & 0.35 \\ 
		\\[-1.8ex]   
		&3&QGC & 0.565 & 0.186 & 0.62 & 0.89 & 4.43 & 0.45 & 4.63 & 0.44 \\ 
		&&BSIM & 0.228 & 0.096 & 0.50 & 0.94 & 4.18 & 0.42 & 4.28 & 0.39 \\ 
		&&BMIM-3 & 0.200 & 0.097 & 0.52 & 0.96 & 4.15 & 0.42 & 4.29 & 0.40 \\ 
		&&BKMR & 0.194 & 0.101 & 0.52 & 0.96 & 4.15 & 0.42 & 4.29 & 0.40 \\ 
		&&BKMR-H*  & 0.277 & 0.118 & 0.38 & 0.80 & 4.22 & 0.44 & 4.28 & 0.37 \\ 
		\hline
	\end{tabular}
\\{\scriptsize *--based on 99 datasets, due to computational instability.}
\end{table}

\section{Analysis of NHANES Case Study}
\label{s:casestudy}
\subsection{Models and Analyses}
\label{ss:casestudy_analyses}

We analyzed the NHANES data using the methods described here in order to compare and contrast the different conclusions that can be drawn under each approach. We considered a broad set of analytic goals: characterizing the overall outcome association of the entire mixture, quantifying the relative contributions of each mixture component, characterizing the outcome associations of each index, and investigating two-way interactions. 

We conducted a QGC analysis (with exposures binned into deciles; $Q=10$),  full BKMR (with componentwise as well as hierarchical variable selection via spike-and-slab priors), a BSIM analysis, and  a  BMIM analysis with three indices (BMIM-3; $M=3$) corresponding to the three exposure groups identified by \cite{gibson2019overview}, within which exposures are expected to act similarly. In the kernel-based methods, exposures were first standardized to have mean 0 and standard deviation 1, and we adopted a Gaussian kernel. All models were adjusted for age (linear and quadratic), sex, and BMI ($<25$, $25-30$,$\geq30$).

\subsection{Results}
\label{ss:casestudy_results}

\subsubsection*{Overall Mixture Association. ~}
All of the methods can estimate an ``overall'' association between mixture and health. However, the interpretations differ somewhat. QGC  quantifies the overall association  through its $\beta$ parameter. This parameter represents  the mean difference in log(LTL) comparing a population to another with all exposures one decile group higher.  We estimate the overall effect to be  0.069 (95\% confidence interval: [0.030, 0.108]). The kernel methods can estimate arbitrary contrasts, so we estimated an analogous ``overall'' mixture-health association, this time comparing a population with all exposures at their 60th percentile versus all exposures set to their 50th percentile. Using this approach, the three kernel methods all produced similar estimates of the overall effect:  the BSIM estimate of this  was 0.034 (95\% credible interval [-0.006, 0.075]), the BMIM-3 estimate was 0.037 (95\%CI [-0.004, 0.077]), and the BKMR estimate was 0.037 (95\% CI [0.005, 0.069]).  There are two likely reason that the estimates of the overall effect vary between QGC and the kernel-based methods.  First, because of the non-linearity in the kernel approaches (as the estimate may depend on the specific quantiles being compared), but also because---despite their similar interpretations---the estimands are slightly different, as QGC compares two quantile \textit{categories} rather than two specific exposure values.

\subsubsection*{Individual Component Contributions. ~}
All of the methods provide a measure of variable or component importance. The interpretation and inference varies between methods.

QGC provides two sets of weights---positive and negative weights---and each represents the proportion of the positive or negative effect that can be attributed to that exposure. Table \ref{tab:NHANESweights} presents estimates of these weights. 
Among the ten exposures with positive weights, Furan1 contributed the most, with a weight 0.18, and PCB99 and Furan4 each had weights of 0.14. Among the negative weights, PCB180 contributed the most, with a weight of 0.29, and the others were no more than 0.16. A key limitation is that the current software provides no estimates of uncertainty for the weights. Hence, inferential statements about the relative size of weights---or even the sign of weights---cannot be readily made.

BKMR does not estimate exposure weights. Instead, we can get a sense of variable importance via posterior inclusion probabilities (PIPs) for the exposures  (Table \ref{tab:NHANESweights}). This is the posterior probability that $\rho_m\neq 0$. The PIP identifies the probability that a particular components contributes to the exposure-response function but does not provide a measure of effect size. This is in sharp contrast to the weights in QGC that identify only the relative size of the association but not statistical certainty. BKMR also identified Furan1 as having by far the strongest signal, with a PIP of 0.88; all other exposures had PIPs below 0.30. PCB180, which had the largest component weight in the QGC analysis, had a PIP of 0.13 indicating weak support for an association with the outcome.  

The BSIM and BMIM-3 estimate PIPs as well as component weights, along with 95\% credible intervals for the weights. The BSIM estimates of the weights for components in the first group (non-dioxin-like PCBs)  were all close to zero (between -0.04 and 0.06; PIPs between 0.14 and 0.17). In contrast, two exposures in this group---PCB99 and PCB180---had among the strongest weights under QGC. Among the 10 other exposures, the signs of the weights matched those of QGC for all but Furan3. Overall, the results were much more similar to those of BKMR, in that Furan1 had by far the strongest weight, with a PIP of 0.81 and an estimated weight of 0.95 (95\% CI [0.00,1.00]); all other estimated weights were far smaller in magnitude and had PIPs of no more than 0.30.  

The BMIM-3 results were similar to those of the BSIM:  Furan1 still had by far the strongest association ($\hat{\theta}_{35}$=0.98, 95\% CI: [-0.12, 1.00]). In index 1, the estimated weight for PCB74 (0.39; 95\% CI [-0.49, 1.00]) was nearly twice that of PCB187 (0.21; 95\% CI [-0.50, 0.99]), indicating the association was twice as strong for PCB74 in the direction of the response curve for index 1---however, index 1 as a whole was very weakly associated with the outcome.

\begin{sidewaystable}[htbp!]
	\centering
	\caption{\label{tab:NHANESweights} Summarizing exposure weights in BKMR and the MIM. For BKMR we report posterior inclusion probabilities (PIPs) for each exposure. For MIM, we report the PIP for the entire index via $\rho$; we also summarize the distribution of weights $\theta$ conditional on $\hat{\rho}\neq0$ (otherwise it is not well defined). Est is the posterior mean standardized to satisfy the constraints; CI is 95\% credible interval.}
	\begin{tabular}{@{\extracolsep{4pt}}rlccccccccccc} 
		\toprule
		& 	&  \multicolumn{2}{c}{QGC} 		 & BKMR & \multicolumn{4}{c}{BSIM ($M$=1)} & \multicolumn{4}{c}{BMIM-3 ($M$=3)} \\
		\cline{3-4}  \cline{5-5}  \cline{6-9}  \cline{10-13} 
		\\[-1.8ex] 
		&		& \multicolumn{2}{c}{Weights}    & 	$\hat{r}_p$      &  $\hat{\rho}_1$  & \multicolumn{3}{c}{$\hat{\theta}_{1l} |\hat{\rho}_1 \neq 0$} &  $\hat{\rho}_m$  & \multicolumn{3}{c}{$\hat{\theta}_{ml} |\hat{\rho}_{m} \neq 0$} \\
		\cline{3-4} \cline{5-5} 		\cline{6-6} 		  \cline{7-9}  \cline{10-10} 		  \cline{11-13} 
		\\[-1.8ex] 
		Group   &  Exposure	 & Pos & Neg  &   PIP       &  PIP & PIP & Est & CI &  PIP & PIP & Est & CI \\ 
		\midrule
		\\[-1.8ex] 
		1& PCB074  &  & 0.10 & 0.11 & 1.00 & 0.17 & 0.04 & (-0.22, 0.58) & 0.60 & 0.37 & 0.39 & (-0.49, 1.00) \\ 
		& PCB099  & 0.14 &  & 0.18 &  & 0.16 & 0.06 & (-0.06, 0.60) &  & 0.39 & 0.60 & (-0.38, 1.00) \\ 
		& PCB138  &  & 0.02 & 0.12 &  & 0.15 & 0.02 & (-0.21, 0.37) &  & 0.33 & 0.29 & (-0.46, 1.00) \\ 
		& PCB153  & 0.11 &  & 0.13 &  & 0.17 & 0.01 & (-0.35, 0.48) &  & 0.40 & 0.38 & (-0.60, 1.00) \\ 
		& PCB170  &  & 0.16 & 0.13 &  & 0.15 & -0.02 & (-0.35, 0.13) &  & 0.37 & 0.25 & (-0.62, 1.00) \\ 
		& PCB180  &  & 0.29 & 0.13 &  & 0.17 & -0.04 & (-0.45, 0.21) &  & 0.41 & 0.28 & (-0.62, 1.00) \\ 
		& PCB187  &  & 0.02 & 0.11 &  & 0.15 & 0.02 & (-0.21, 0.41) &  & 0.34 & 0.21 & (-0.5, 0.99) \\ 
		& PCB194  & 0.08 &  & 0.09 &  & 0.14 & -0.01 & (-0.32, 0.20) &  & 0.36 & 0.28 & (-0.53, 1.00) \\ 
		\\[-1.8ex]     
		2& PCB126  & 0.09 &  & 0.09 &  & 0.16 & 0.07 & (-0.03, 0.67) & 0.44 & 0.51 & 0.46 & (-0.18, 1.00) \\ 
		& PCB169  & 0.10 &  & 0.18 &  & 0.25 & 0.13 & (-0.07, 0.85) &  & 0.79 & 0.89 & (0.00, 1.00) \\ 
		\\[-1.8ex]     
		3& PCB118  & 0.10 &  & 0.11 &  & 0.30 & 0.24 & (-0.02, 0.91) & 0.97 & 0.25 & 0.15 & (-0.11, 0.94) \\ 
		& Dioxin1 & 0.05 &  & 0.10 &  & 0.13 & 0.01 & (-0.20, 0.30) &  & 0.19 & -0.02 & (-0.48, 0.37) \\ 
		& Dioxin2 &  & 0.11 & 0.09 &  & 0.14 & -0.03 & (-0.37, 0.07) &  & 0.23 & -0.04 & (-0.61, 0.49) \\ 
		& Dioxin3 &  & 0.16 & 0.07 &  & 0.11 & -0.02 & (-0.28, 0.02) &  & 0.17 & 0.00 & (-0.36, 0.41) \\ 
		& Furan1  & 0.18 &  & 0.88 &  & 0.81 & 0.95 & (0.00, 1.00) &  & 0.82 & 0.98 & (-0.12, 1.00) \\ 
		& Furan2  & 0.02 &  & 0.10 &  & 0.16 & 0.05 & (-0.13, 0.47) &  & 0.21 & 0.06 & (-0.44, 0.76) \\ 
		& Furan3  &  & 0.14 & 0.11 &  & 0.15 & 0.05 & (-0.05, 0.46) &  & 0.24 & 0.08 & (-0.38, 0.87) \\ 
		& Furan4  & 0.14 &  & 0.21 &  & 0.11 & 0.01 & (-0.11, 0.25) &  & 0.22 & 0.11 & (-0.3, 0.94) \\ 
		\bottomrule
	\end{tabular}
\end{sidewaystable}

\begin{figure}[htbp!]
	\includegraphics[width=0.99\linewidth]{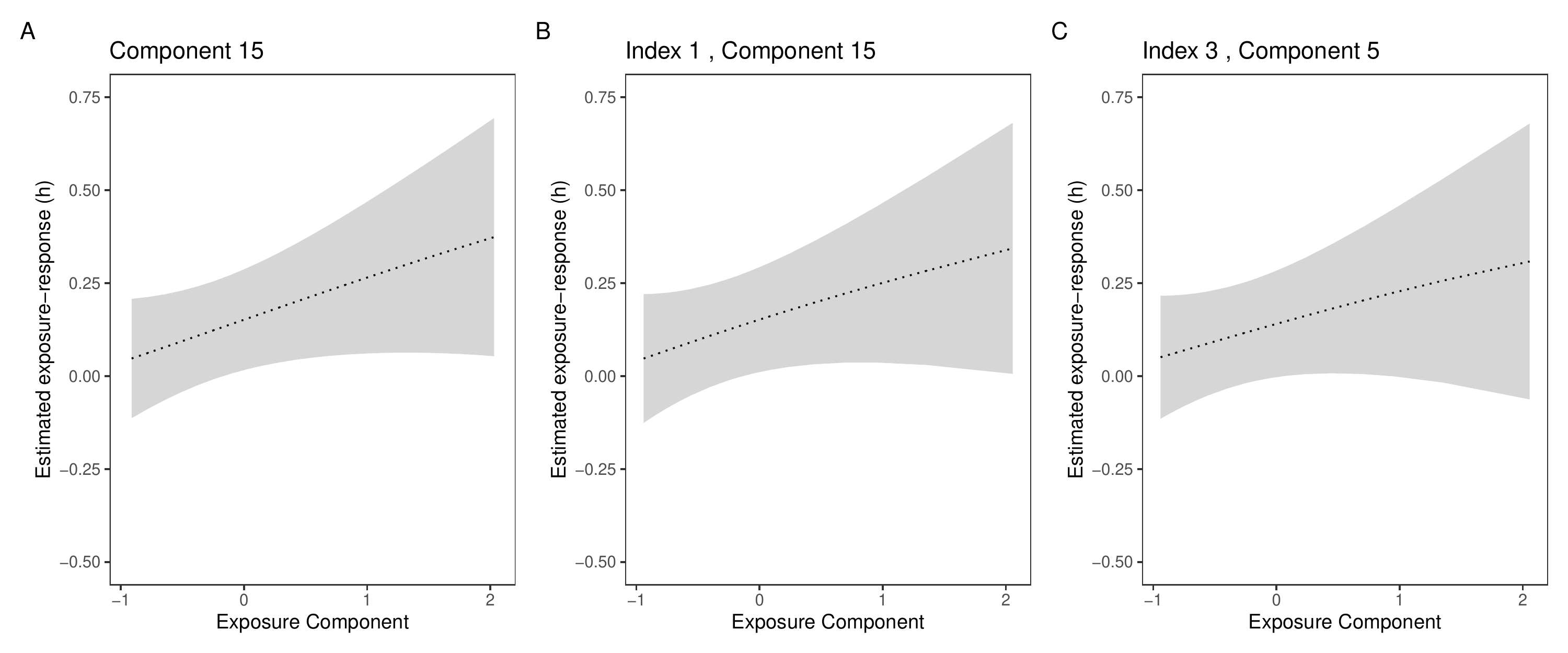} 
	\caption{Exposure response curve for Furan1 in NHANES analysis. Panel A shows fitted curve using the {\tt bkmr} package in R; B is based on single index model; C is based on 3-index model.}
	\label{fig:pp15}
\end{figure}

As Furan1 was identified as the strongest mixture component by each of the kernel methods, we plot the corresponding estimated exposure-response curves (with all other exposures set to their medians) under each of these approaches in Figure \ref{fig:pp15}. The three plots are nearly identical, indicating that BKMR, the BSIM and the BMIM-3 lead to roughly the same conclusions for the association between Furan1 and the outcome.

\subsubsection*{Indexwise Exposure-Response Curves. ~}

\begin{figure}[htbp!]
	\centering
	\subfloat[Bayesian single index model ($M$=1; $L_1$=18)]{
		\includegraphics[width=0.4\linewidth]{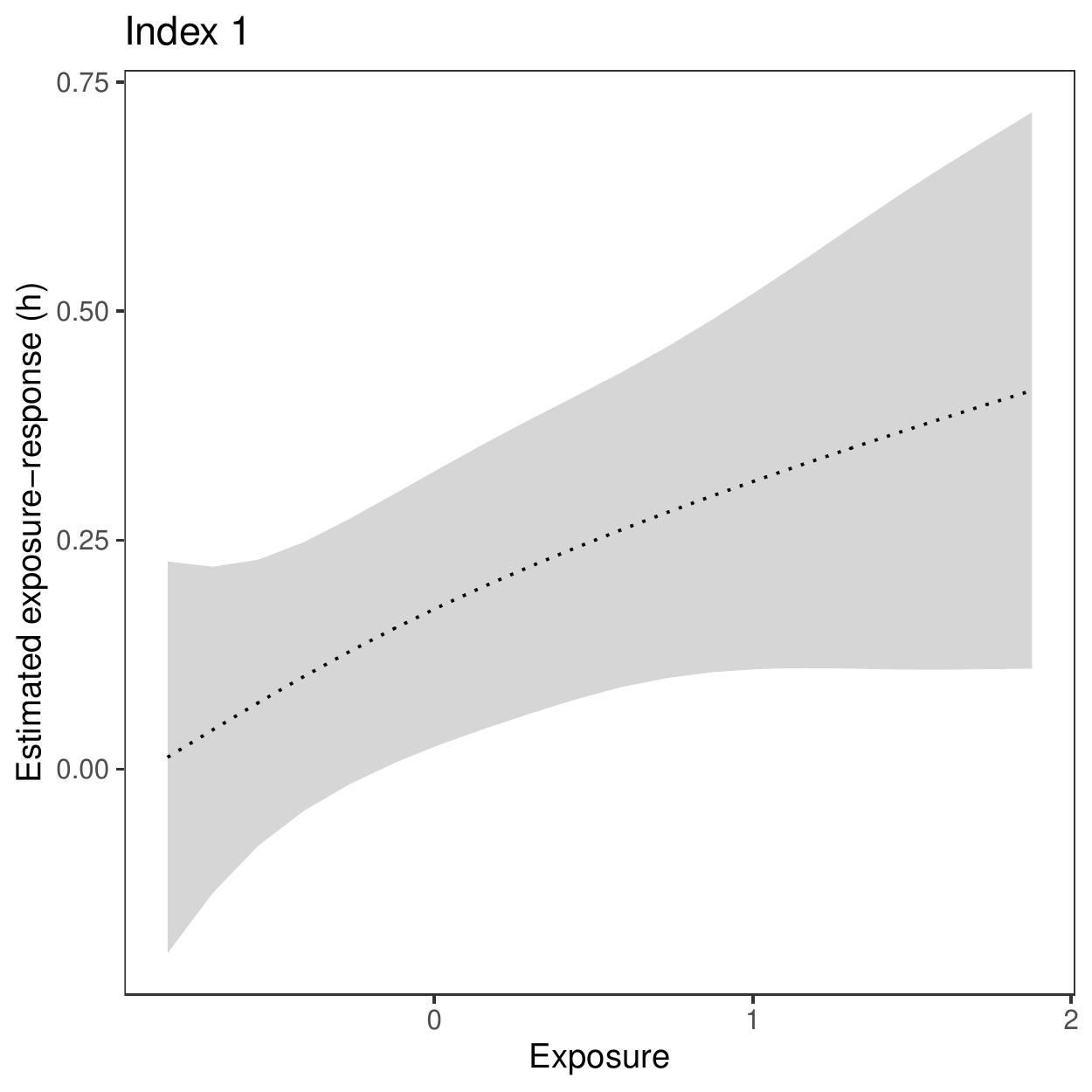}} \\
	\centering
	\subfloat[Bayesian multiple index model ($M$=3; $L_1$=8, $L_2$=2, $L_3$=8)]{
		\includegraphics[width=\linewidth]{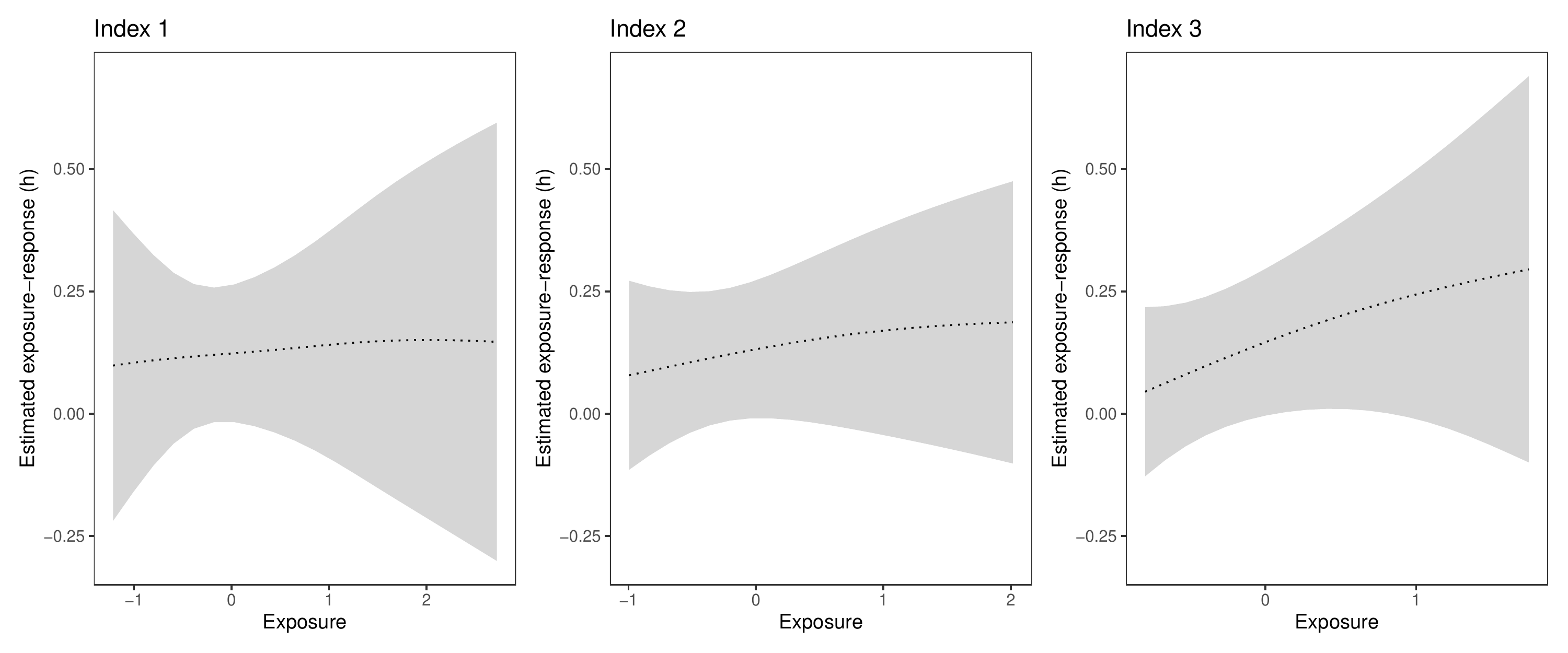}}
	\caption{Estimated index-wise response curves from NHANES analysis. The top panel shows the exposure-response function for the BSIM. The bottom panel shows the exposure-response functions for each of the three indices in the BMIM-3. For the BMIM-3, index exposure-response function is plotted with the other indices fixed at the median value. The plot also shows 95\% credible intervals.}
	\label{fig:pp_ind}
\end{figure}

One of the advantages of the proposed index models is that they naturally facilitate studying response curves at the index level, rather than reporting $P=18$ individual exposure-response curves, or increasing every exposure simultaneously. Under the BSIM, we visualize the entire index-wise estimated response curve in Figure \ref{fig:pp_ind} (a), which appears just slightly sub-linear. Under the BMIM-3, the three estimated index-wise exposure response curves are not radically different from that of the BSIM (panel b), with each increasing sublinearly. Specifically, the estimated response curve for index 1 is close to null (flat) and has the widest credible intervals; that of index 2 corresponds to a slightly stronger association, and that of index 3 is the strongest.  Note in particular the similarity between the response curve for index 3 and the component-wise curve for Furan1. Matching this observation, the index-level PIPs (for $\rho_m$) were correspondingly low for the first two indices (0.60 and 0.44), whereas that of index 3 was 0.97.

\subsubsection*{Interactions. ~}
\begin{sidewaysfigure}[htbp!]
	\subfloat[BKMR ($M$=$P$=18)]{
		\includegraphics[width=0.6\linewidth]{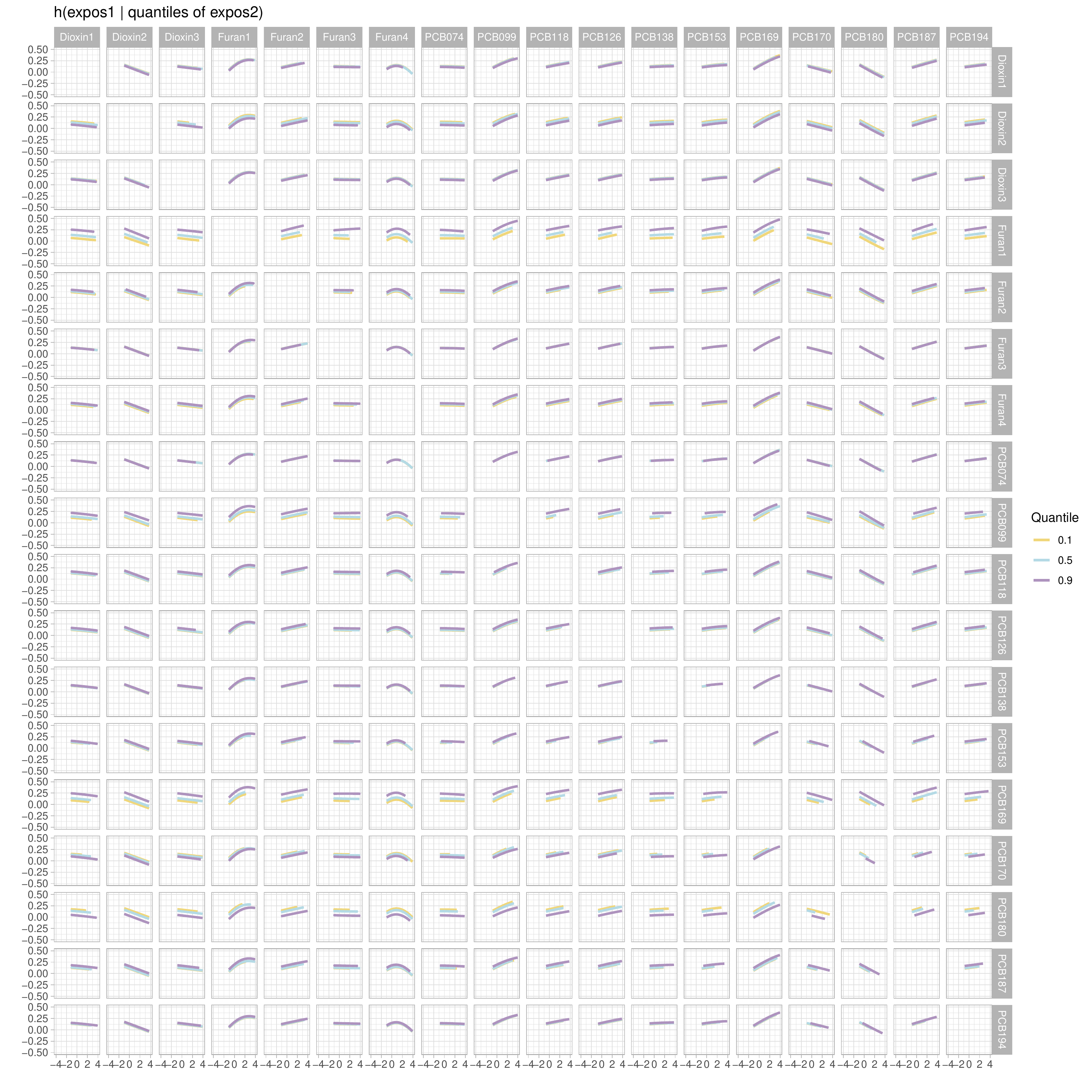}} 
	\subfloat[Bayesian multiple index model ($M$=3; $L_1$=8, $L_2$=2, $L_3$=8)]{
		\includegraphics[width=0.4\linewidth]{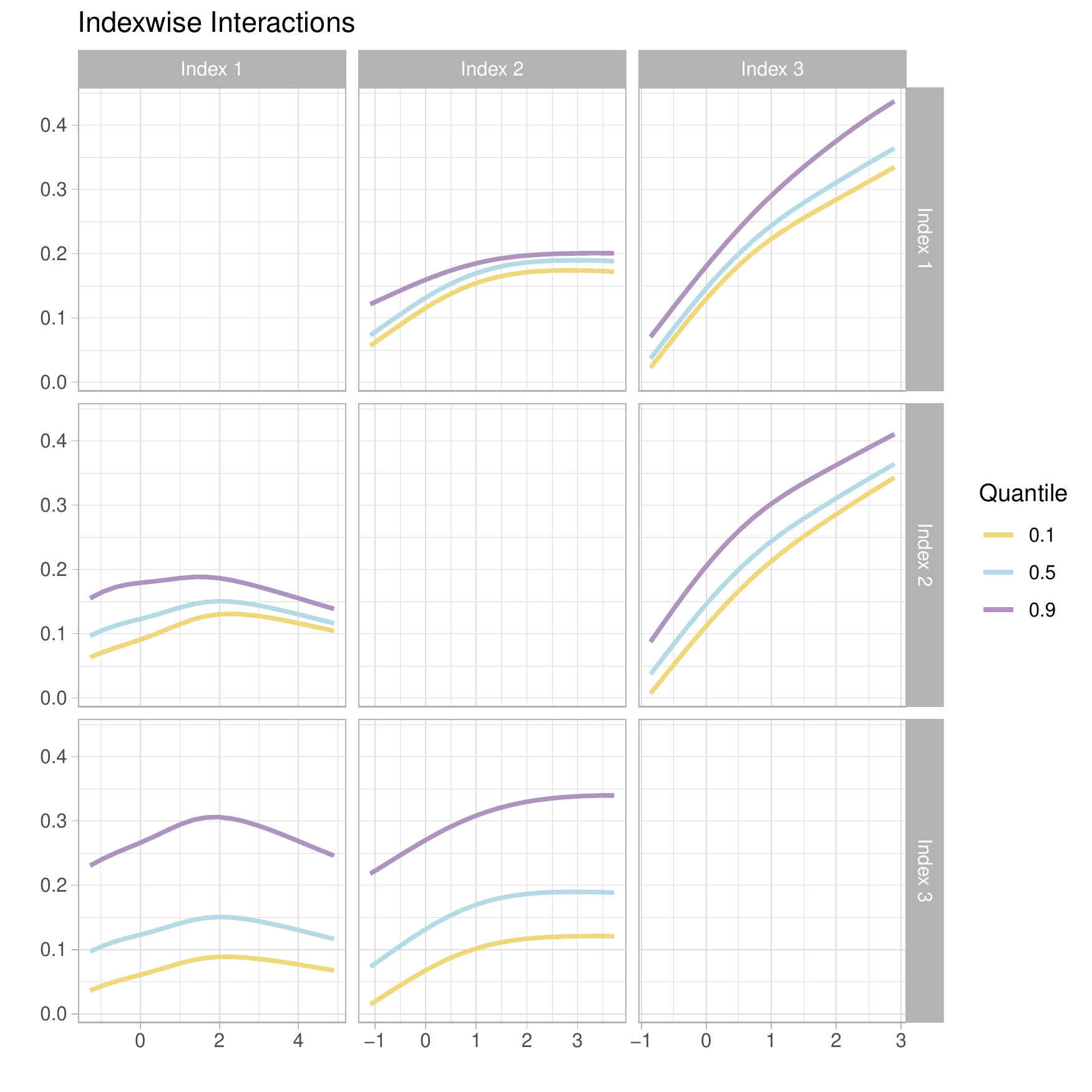}}
	\caption{Visualizing two-way interactions under BKMR (left) and the BMIM-3 (right) in the NHANES analysis. Each panel shows the exposure-response relation for one exposure (panel [a]) or one index (panel [b]) with all other exposures or indices fixed at a given percentile (0.1, 0.5, or 0.9).  Parallel lines indicate no interaction between components or indices while deviation indicates interactions. Color version can be found online in the electronic version of the article.}
	\label{fig:interactionsNHANES}
\end{sidewaysfigure}

A feature of the BMIM and BKMR approaches is that we can consider interactions, although we consider a different set of interactions in each. Under BKMR, we compare fitted exposure response curves for $x_{j}$ when $x_{j'}$ is set to its $10^{th}$, $50^{th}$ or $90^{th}$ percentile (and all others to their medians). We display these for all pairs $j\neq j'$ in Figure \ref{fig:interactionsNHANES} (a), although it is not straightforward to interpret the 306 ($P\times [P-1]$) resulting plots. By contrast the BMIM-3 framework allows one to characterize interactions between entire \textit{indices}, so that rather than scanning 306 plots, we need only investigate 6 ($M\times [M-1]$). Moreover, as indexwise plots more naturally characterize how exposures within an index vary jointly, they are less prone to the inevitable sparsity issues that arise in the 306 component-wise interaction plots. As such, we plot these two-way indexwise interactions in panel (b), and there appears to be some indication of interaction between index 1 and the others, although the evidence is weak relative to the level of uncertainty. 

\subsubsection*{Model Fit. ~} 
We also conducted 5-fold cross validation in order to compare model fit via MSE, as in the simulations.  Of the four models, QGC had the highest MSE (0.790), and BKMR had the lowest (0.774). From a statistical perspective, the MSEs for BSIM and BMIM-3 were effectively equal to that of BKMR (less than 0.3\% higher). As the results of the kernel approaches identified one strong association, we also fit BKMR with hierarchical variable selection, but it fit no better than the standard BKMR (0.774). Practically speaking, the simpler index models were able to achieve virtually the same fit as the more complex BKMR, all while simplifying the burden of interpretation substantially.

\section{Discussion}
\label{s:discussion}

The proposed BMIM framework represents a spectrum of models, with linear index models and BKMR occupying either end of the spectrum. At present it is common in environmental mixture studies to adopt one of these two extremes.  A key goal of this paper is to increase the range of options available to an analyst and unify the two seemingly unrelated methods. 
By bridging the gap between the two,  this framework gives analysts the freedom to select an analysis strategy with an appropriate balance of flexibility and interpretability. 
At its most parsimonious, the BMIM extends the standard linear index models to allow for variable selection and non-linear relations. When $M>1$,  and in the absence of interactions, it can also allow for additive index models, which have not been explored in the environmental health literature. Even when a single multi-exposure index is believed to be appropriate, the BMIM allows one to examine interactions between this multi-exposure index and one or more individual covariates like age or a measure of socioeconomic status (SES), e.g., $y_i=h^3\left(\mathbf{x_{i}}^T \boldsymbol{\theta}_1,age_i,SES_i\right) +\mathbf{z}_i^T\boldsymbol{\gamma}+\epsilon_i$. 

 Unlike fully non-parametric approaches, the BMIM framework allows one to incorporate and investigate biologically plausible mechanisms. If, say, a set of compounds is hypothesized to bind to the same receptor, an index structure might allow one to easily encode this prior knowledge. In toxicology, it is common to construct multi-pollutant indices based on relative potency factors (e.g. toxic equivalency factors) derived from laboratory experiments \citep{mitro2016cross}. The BMIM allows one to incorporate such indices while allowing for a non-linear relationship and allowing one to investigate possible interactions among indices without fully specifying a parametric relationship.
 
 In the NHANES case study, some commonalities and a few interesting differences emerged across methods.  All methods weighted the Furan1 association highly.  The BSIM, BMIM-3, and BKMR all indicated it had the strongest association with the outcome;  QGC estimated it to have the largest positive weight.  An interesting difference is that  QGC reported an even larger negative weight for PCB180, whereas the response-surface and Bayesian index methods did not weight this exposure highly (based on either posterior inclusion probabilities or component weights).  This occurs because while the estimated coefficient for this exposure in the underlying linear model is the largest in magnitude, it is not large relative to the uncertainty of the estimate.  The Bayesian methods reflect this uncertainty, and therefore the strength of evidence of an association with a particular exposure, whereas the convention for existing index models in the literature has been to not report these uncertainties. Our results suggest this is worth doing even if one opts for an existing frequentist index method.

 A feature of the proposed framework is the ability to estimate indexwise associations. From a substantive perspective, this   more naturally captures how exposures within an index vary jointly---rather than artificially setting them all to some fixed quantile simultaneously.  One might even estimate an overall mixture effect in the same way, say by contrasting all indices set to two quantiles. From a statistical perspective, this can also help alleviate sparsity  issues. For example it is typical to summarize interactions with BKMR by plotting an estimated exposure response curve at some fixed level of another exposure and at median levels of all others. This might be reasonable for a subset of the exposures, but one quickly encounters sparsity when examining the 153 different pairwise comparisons in the NHANES data. 
 This is exacerbated when one considers higher order interactions.  By instead basing inference on composites of multiple exposures, the BMIM is less reliant on such fine-grained comparisons.

	
	\backmatter
	
	\vspace*{-20pt}
	\section*{Acknowledgements}
	This research was supported by NIH grants ES028800, ES028811, ES028688, and ES000002. This research was also supported by USEPA grants RD-835872-01 and RD-839278.  Its contents are solely the responsibility of the grantee and do not necessarily represent the official views of the USEPA.  Further, USEPA does not endorse the purchase of any commercial products or services mentioned in the publication.
\vspace*{-8pt}

%
	\section*{Supporting Information}
Supplementary material is available online. Software is available at {\tt github.com/glenmcgee/BMIM}.\vspace*{-8pt}

	
	%
	\bibliographystyle{biom} 
	\bibliography{McGee_Bibliography.bib}

	\label{lastpage}
	
\end{document}